\documentclass[pra,preprint,showkeys,showpacs,amsmath,amssymb]{revtex4}
\usepackage{bm}
\usepackage{graphics}

\begin{document}
\title{Mixed-isotope Bose-Einstein condensates in Rubidium}

\author{Valery S. Shchesnovich$^{1}$ }
\email{valery@ift.unesp.br}
\author{Anatoly M. Kamchatnov$^{1,2}$}
\email{kamch@isan.troitsk.ru}
\author{Roberto A. Kraenkel$^1$}
\email{kraenkel@ift.unesp.br}

\affiliation{
$^1$Instituto de F\'{\i}sica Te\'{o}rica, Universidade Estadual Paulista-UNESP,
Rua Pamplona 145, 01405-900 S\~{a}o Paulo, Brazil\\
$^2$Institute of Spectroscopy, Russian Academy of Sciences, Troitsk 142190,
Moscow Region, Russia}

\date{\today}

\begin{abstract}
We consider  the ground state properties of mixed Bose-Einstein condensates
of ${}^{87}$Rb and ${}^{85}$Rb atoms in the isotropic pancake trap, for both
signs of the interspecies scattering length. In the case of repulsive
interspecies interaction, there are the axially-symmetric and
symmetry-breaking ground states. The threshold for the symmetry breaking
transition, which is related to appearance of a zero dipole-mode, is found
numerically. For attractive interspecies interactions, the two condensates
assume symmetric ground states for the numbers of atoms up to the collapse
instability of the mixture.
\end{abstract}



\maketitle

\section{Introduction}
\label{I}

Bose-Einstein condensation (BEC) in mixtures of trapped quantum gases has
become an exciting field of study. First experimental observation of the
two-species BEC was realized using two different spin states of ${}^{87}$Rb
\cite{expr_twoRb87a}. The two overlapping condensates of ${}^{87}$Rb in the
spin states $|F=1,m=-1\rangle$ and $|F=2, m=2\rangle$ were created via nearly
lossless sympathetic cooling of the atoms in the state $|2, 2\rangle$  by
thermal contact with the atoms in the $|1,-1\rangle$-state. Also, the
double-condensate system of ${}^{87}$Rb in the spin states $|1,-1\rangle$ and
$|2, 1\rangle$ was created from the single condensate in the
$|1,-1\rangle$-state by driving a two-photon transition \cite{expr_twoRb87b}.
In the subsequent evolution after creation, the condensates underwent complex
relative motions preserving the total density profile. The motions quickly
damped out and the condensates assumed a steady state with a non-negligible
(and adjustable) overlap region. These experiments started a series of works
devoted to experimental and theoretical study of BEC in mixtures. For
instance, superposition of the spinor condensates of ${}^{23}$Na led to
observation of weakly miscible and immiscible superfluids \cite{exprc} and
occurrence of the metastable states \cite{exprd}. Interaction between two
condensates of different spin states of ${}^{87}$Rb in the displaced traps
was observed in the center-of-mass oscillations \cite{expre}. Successful
attempts to cool fermion gases to the quantum degeneracy regime by using the
boson-fermion mixtures were also reported. First such mixture was achieved by
using the two species of Li, the bosonic ${}^7$Li and fermionic ${}^6$Li
\cite{Li67a,Li67b}. More recently, the experiments on mixtures of different
atomic species were performed. Both boson-boson and boson-fermion pairs were
cooled. The two species BEC of ${}^{41}$K and ${}^{87}$Rb \cite{K41Rb87} and
boson-fermion mixtures ${}^{87}$Rb--${}^{40}$K \cite{K40Rb87a,K40Rb87b} and
${}^{23}$Na--${}^6$Li \cite{Li6Na23} were achieved.

A promising combination for  obtaining the two-species BEC potentially rich
in new phenomena is the mixture of two isotopes of Rubidium: ${}^{85}$Rb and
${}^{87}$Rb. There is a long standing interest in obtaining this BEC mixture,
which goes back to Ref.~\cite{prospects}, where the feasibility of achieving
such two-species BEC  was established. It was suggested that condensation of
the two isotopes of Rubidium can be achieved via the sympathetic cooling of
certain hyperfine states which exhibit low inelastic collision rates.
Moreover, the possibility of employing the Feshbach resonance for control
over the scattering length was stressed. The optimal combination was found to
be the mixture of the spin states $|2,-2\rangle_{85}$ and
$|1,-1\rangle_{87}$, because the scattering length {\it between} the isotopes
can be controlled. The sympathetic cooling of the ${}^{85}$Rb condensate by
thermal contact with the ${}^{87}$Rb condensate was subsequently
experimentally demonstrated \cite{sympath_cool}. Up to $10^6$ atoms  of the
${}^{85}$Rb isotope were cooled via elastic collisions with a large reservoir
($10^9$ atoms) of ${}^{87}$Rb. The stable condensate of the ${}^{85}$Rb
isotope was also created by using the Feshbach resonance to reverse the sign
of the scattering length from negative to positive \cite{Rb85expr}. In this
way, the long-living condensates with up to $10^4$ atoms of ${}^{85}$Rb in
the spin state $|2,-2\rangle$ were produced.

One of the principal  advantages of using the Rubidium isotopes is that their
interspecies and intraspecies  scattering lengths are known with a good
precision \cite{prospects}, thus  theoretical predictions can be compared
with the experiment. In particular, the scattering lengths of the ${}^{87}$Rb
isotope  and between the two isotopes are positive, while the scattering
length of the ${}^{85}$Rb isotope is negative.

Efficient interspecies  thermalization crucially depends on the interspecies
scattering length and the overlap region of the species. It is known that the
spatial separation may take place depending on the values of the scattering
lengths. If all atomic interactions in the mixture are repulsive, the
following simple criterion for the spatial separation of two BECs in a box
\cite{crit_repuls} is known: if the mutual repulsion is large enough, namely
$G_{12}>\sqrt{G_{11}G_{22}}$ (where $G_{ij}$ is the interaction coefficient),
the condensates separate to lower the energy. The symmetry breaking point of
view on the ground state in the mixture of condensates was developed in
Refs.~\cite{symbr1,symbr2,symm_break,symm_br_pan,sym_asym_trans}. For
instance, by taking equal number of atoms in the two species, the
symmetry-preserving vs. symmetry-breaking phase diagram was obtained in
Ref.~\cite{symm_break}. Existence of the metastable states in the BEC
mixtures  was argued also on the basis of the Bogoliubov excitation spectra
in Ref.~\cite{spectrum}, where both signs of the interspecies scattering
length were considered (for the repulsive intraspecies interactions). In
Ref.~\cite{variat} the two-species condensate with coinciding positive or
negative interspecies scattering lengths and equal numbers of atoms in the
species were considered within a variational approach. However, the results
of the latter work do not apply to the BEC mixture of the two isotopes of
Rubidium, where, first of all, the interspecies scattering lengths have
different signs. Finally, the  collapse of a two-component BEC in the
spherically symmetric trap was  numerically studied in Ref.~\cite{multicoll},
where all possible combinations of signs of the atomic interactions for the
two species were considered. It was found that, depending on the interaction
coefficients, either one or both components  may experience collapse.

In the related theoretical studies of boson-fermion mixtures
\cite{TF1,GPETF,Mean-Field,Three-Fluid,CollapseBF,FullBF} all possible signs
of the boson and boson-fermion $s$-wave scattering lengths were considered
(due to the strong $s$-wave scattering between bosons and fermions the
$p$-wave contribution to the interspecies interaction is neglected, see
Refs.~\cite{Mean-Field,FullBF}). This reflects the fact that in the
experiments on boson-fermion mixtures various combinations of the signs are
possible, for instance, the two-isotope mixture of  Lithium of
Ref.~\cite{Li67a} had attractive boson and repulsive boson-fermion
interactions, while in Ref.~\cite{Li67b} the same species were used in the
different angular momentum states with the repulsive atomic interactions.
Though the governing equations for the boson-fermion mixture are different
from those for the two-boson BEC, the predicted effects, such as the phase
separation \cite{TF1,FullBF,Three-Fluid} and collapse
\cite{CollapseBF,FullBF}, have similar features.  In Ref.~\cite{FullBF} the
comprehensive analysis of the properties of boson-fermion mixtures for all
possible signs of the boson and boson-fermion $s$-wave scattering lengths is
given. We will make connections to the results of the latter work in the
following sections.

In the present paper we study two-species BEC in a pancake trap, for the
numbers of atoms below the collapse instability. Our main goal is to
understand  the ground state of the two-species BEC mixture comprised of the
${}^{85}$Rb and ${}^{87}$Rb isotopes, with the atoms being in the optimal
spin states, $|2,-2\rangle_{85}$ and $|1,-1\rangle_{87}$. We consider both
the attractive and the repulsive interspecies interactions for fixed
(default) intraspecies interactions with the scattering lengths: $-412.5$
a.u. ($-21.8$ nm) for $|2,-2\rangle_{85}$ and $107.5$ a.u. ($5.7$ nm) for
$|1,-1\rangle_{87}$---the averages of those given in Ref.~\cite{prospects}.
In section \ref{II} we introduce the two-dimensional model describing the
two-species BEC mixture in a pancake trap and discuss the domain of its
applicability. Then, we present the numerically found ground states in the
BEC mixture of the two isotopes of Rubidium for the repulsive as well as
attractive interspecies interactions, sections \ref{IIIa} and \ref{IIIb},
respectively. The concluding section \ref{IV} contains brief summary of the
results. Detailed derivation of the two-dimensional model is placed in
appendix~\ref{appendA}, while details of the stability analysis of the
axially symmetric ground states are given in appendix~\ref{appendB}.

\section{Two-dimensional model for the pancake trap}
\label{II}

We consider two-species BEC mixture in the isotropic pancake trap for
arbitrary intraspecies and interspecies scattering lengths. The
Gross-Pitaevskii equations for the two-species BEC have the following form
\cite{magn-prop}:
\begin{subequations}
\label{Eq1ab}
\begin{eqnarray}
i\hbar \partial_t \Psi_1 &=& -\frac{\hbar^2}{2m_1} \nabla^2\Psi_1 +
V_1(\bm{r})\Psi_1 + (G_{11}|\Psi_1|^2 + G_{12}|\Psi_2|^2)\Psi_1,\\
i\hbar \partial_t \Psi_2 &=& -\frac{\hbar^2}{2m_2} \nabla^2\Psi_2 +
V_2(\bm{r})\Psi_2 + (G_{22}|\Psi_2|^2 + G_{12}|\Psi_1|^2)\Psi_2,
\end{eqnarray}
\end{subequations}
where $\Psi_1(\bm{r},t)$ and $\Psi_2(\bm{r},t)$ are the order parameters of
the two species, while the interaction coefficients are given as $G_{11} =
4\pi\hbar^2 a_1/m_1$, $G_{22} = 4\pi\hbar^2 a_2/m_2$, and $G_{12} =
2\pi\hbar^2a_{12}/M$, with $a_1$, $a_2$ and $a_{12}$ being the intraspecies
and  interspecies scattering lengths, respectively. Here $M$ denotes the
reduced mass: $M = m_1m_2/(m_1+m_2)$. In our case, for the two isotopes of
Rubidium, we can neglect the mass difference between the isotopes and take $m
= m_1 = m_2$. We consider the parabolic pancake trap,
\begin{equation}
V_j = \frac{m\omega_{j,z}^2}{2}z^2 +
\frac{m\omega_{j,\perp}^2}{2}r_\perp^2, \quad
j=1,2, \label{Eq2}
\end{equation}
with strong confinement in  the $z$-direction: $\gamma \equiv
\omega_z/\omega_\perp \gg1$ (by a simple phase transformation the possible
difference of the zero-point energies for the two species in the trap can be
scaled away from system (\ref{Eq1ab})). The difference in the magnetic trap
frequencies felt by the two species is caused by the difference of the Lande
magnetic factors for the two isotopes \cite{magn-prop}: $g(|2,-2\rangle_{85})
= -1/6$ and $g(|1,-1\rangle_{87}) = -1/4$. The corresponding magnetic moments
measured in the Bohr magnetons are given as follows: $\mu_{85}
=g(|2,-2\rangle_{85})m_{85} = 1/3$ and $\mu_{85} = g(|1,-1\rangle_{87})m_{87}
=1/4$. Hence, the ratio of the squared trap frequencies is
$\omega^2_{87}/\omega^2_{85} = \mu_{87}/\mu_{85} = 3/4$, where $\omega_{85}$
and $\omega_{87}$ stand for the frequencies experienced by the respective
isotopes. From now on, the indices 1 and 2 will correspond to the isotopes
${}^{85}$Rb and ${}^{87}$Rb, respectively.

For not too large numbers of atoms the three-dimensional system (\ref{Eq1ab})
can be reduced to a system of two-dimensional equations in the pancake
coordinates $\bm{r}_\perp = (x,y)$, while the order parameter in the
$z$-direction is fixed and given by the Gaussian. Indeed, the motion in the
$z$-direction is quantized under the condition that the energy contribution
from the nonlinearity is much less than the difference between the energy
levels of the trap:
\begin{equation}
\frac{|G_{j1}|N_1}{d_{1,z}d^2_{1,\perp}} \ll
\hbar\omega_{1,z},\quad \frac{|G_{j2}|N_2}{d_{2,z}d^2_{2,\perp}}
\ll \hbar\omega_{2,z},\quad j=1,2.
\label{Eq3}\end{equation}
 Here we have estimated the order parameter as $|\Psi_j|^2 \sim
N_j/(d_{j,z}d_{j,\perp}^2)$, $j=1,2$, with introduction of the effective
sizes of  the condensates in the pancake plane ($d_{j,\perp}$) and in the
$z$-direction ($d_{j,z}$). Under  condition (\ref{Eq3}), the $z$-sizes of the
condensates are given by the trap size: $d_{j,z} = a_{j,z}$, with $a_{j,z}$
being the respective oscillator length in the $z$-direction (see formula
(\ref{Eq4})). Whereas their sizes in the pancake plane must be determined
from the solution to the resulting two-dimensional system (system
(\ref{Eq6ab}) below). We will reformulate  condition (\ref{Eq3}) in a more
convenient form below. More detailed analysis of the two-dimensional
approximation is placed in appendix~\ref{appendA}.

Under condition (\ref{Eq3}) the order parameter $\Psi_j$ is approximated as a
product of the Gaussian wave function in the $z$-direction and a wave
function describing the transverse shape:
\begin{equation}
\Psi_j = e^{-i\omega_{j,z}t/2} f_j(z)\Phi_j(\bm{r}_\perp,t), \quad
f_j\equiv\pi^{-1/4}a_{j,z}^{-1/2}\exp\left(-\frac{z^2}{2a^2_{j,z}}\right),
\quad a_{j,z} \equiv
\left(\frac{\hbar}{m\omega_{j,z}}\right)^{1/2}.\label{Eq4}
\end{equation}
The Gaussian is the ground state wave-function of the linear part of the
r.h.s. in system (\ref{Eq1ab}) which correspond to the quantum motion in the
$z$-direction: $H_{j,z}\equiv -\hbar^2/(2m)\partial^2_z +
m\omega^2_{j,z}z^2/2$, with $H_{j,z}f_j(z) = (\hbar\omega_{j,z}/2)f_j(z)$.
Substitution of  expression (\ref{Eq4}) in system (\ref{Eq1ab}),
multiplication of the equation for $\Psi_j$ by $f_j(z)$ and integration over
$z$ results in the approximate two-dimensional system (see also equations
(\ref{B3a})-(\ref{B3b}) from appendix~\ref{appendA}). To write it down in a
form convenient for numerical calculations, let us introduce the
dimensionless variables:
\begin{equation}
\bm{\rho} = \frac{\bm{r}_\perp }{a_\perp}, \quad a_\perp \equiv
\left(\frac{\hbar}{m\omega_{\perp}}\right)^{1/2},\quad T =
\frac{\omega_\perp}{2}t,\quad \psi_j = {a_\perp}\Phi_j,\quad
j=1,2. \label{Eq5}
\end{equation}
Here we have defined the frequency $\omega_\perp$ as
$\omega^2_\perp=(\omega^2_{1,\perp}+\omega^2_{2,\perp})/2$, where
$\omega_{j,\perp}$, $j=1,2$, are the trap frequencies in the
pancake plane experienced by the two isotopes. Then the
dimensionless  approximate 2D system reads:
\begin{subequations}
\label{Eq6ab}
\begin{eqnarray}
i\partial_T \psi_1 &=& -\nabla^2_\perp\psi_1 +
\lambda^2_1\rho^2\psi_1 + (g_{11}|\psi_1|^2 + g_{12}|\psi_2|^2)\psi_1,
\label{Eq6a}\\
i \partial_T \psi_2 &=& -\nabla^2_\perp\psi_2 + \lambda^2_2\rho^2\psi_2 +
(g_{22}|\psi_2|^2 + g_{12}|\psi_1|^2)\psi_2,\label{Eq6b}
\end{eqnarray}
\end{subequations}
where $\rho = |\bm{\rho}|$,
\begin{equation}
g_{11} = \frac{4\sqrt{2\pi}a_{1}}{a_{1,z}},\quad g_{22} =
 \frac{4\sqrt{2\pi}a_{2}}{a_{2,z}},\quad g_{12} =
\frac{8\sqrt{\pi}a_{12}}{(a_{1,z}^2+a_{2,z}^2)^{1/2}},\quad
\lambda_1 = \frac{\omega_{1,\perp}}{\omega_\perp},\quad \lambda_2
= \frac{\omega_{2,\perp}}{\omega_\perp}. \label{Eq7}
\end{equation}
Using the relation $\omega^2_{2}/\omega^2_{1} = 3/4$ for the
Rubidium isotopes in the spin states $|2,-2\rangle_{85}$ and
$|1,-1\rangle_{87}$, we obtain: $\lambda^2_1 = 8/7$ and
$\lambda^2_2 = 6/7$.

The pancake trap sizes in the experiments on BEC have different values. To
set a reference for discussion, in the calculations below we assume the
$z$-size of the trap to be 10$\mu$m, i.e., we set $a_{1,z} = 10\mu$m
($a_{2,z} = 2.03^{-1/4}a_{1,z}=0.84a_{1,z}$). This results in the following
values for the interaction coefficients in the mixture of the two isotopes of
Rubidium: $g_{11} = -0.0219$, $g_{22}= 0.0068$, and $g_{12} = 0.012$. For a
different trap size the interaction coefficients will change. However, the
quantities $g_{11}N_1$, $g_{22}N_2$, $g_{12}N_2$, and $g_{12}N_1$, computed
on a solution to system (\ref{Eq6ab}), will remain invariant under the
variation of the trap size. Thus, a different trap size $a_z$ will result in
a similar solution but for the appropriately scaled numbers of atoms. We will
return to this point below.

Let us now reformulate condition (\ref{Eq3}) in a form more convenient for
verification. Scaling the  sizes of the condensates in the pancake plane by
the respective trap length, $d_{j,\perp} = R_ja_\perp$, we obtain the
equivalent conditions in the form of the bounds on the numbers of atoms:
\begin{equation}
N_1 \ll \gamma\frac{R^2_1}{4\pi}\mathrm{min}
\left(\frac{a_{z}}{|a_1|},\frac{a_{z}}{|a_{12}|}\right), \quad
N_2\ll\gamma\frac{R^2_2}{4\pi}\mathrm{min}
\left(\frac{a_{z}}{|a_{2}|},\frac{a_{z}}{|a_{12}|}\right),
 \label{Eq8}\end{equation}
where $a_z$ denotes both $a_{1,z}$ and $a_{2,z}$, since they have close
values, and  $\gamma\gg1$ ($\gamma = \omega_z/\omega_\perp =
a_\perp^2/a_z^2$). The sizes $R_1$ and $R_2$ of the two condensates must be
determined from the solution of system (\ref{Eq6ab}). For instance, for the
pancake trap with $a_z = 10\mu$m, using the values of the scattering lengths
from section~\ref{I}  for the ${}^{85}$Rb--${}^{87}$Rb mixture, we obtain the
following bounds: $N_j \ll 10^2\gamma R_j^2$, $j=1,2$. The actual bounds on
the numbers of atoms are thus determined by the trap anisotropy $\gamma$. For
example, if $\gamma = 100$ (i.e., $a_\perp =10 a_z$) we have $N_j \ll 10^4
R_j^2$.

There is the critical number of atoms $N_c$ such that the
${}^{85}$Rb condensate, in the absence of the other isotope, is
unstable with respect to collapse for $N_{85}>N_c$. By setting
$g_{12} = 0$ in the two-dimensional system (\ref{Eq6ab}), we
obtain the following expression for the critical number of the
${}^{85}$Rb atoms necessary for collapse (in the absence of the
other isotope):
\begin{equation}
N_{c} = \frac{2\pi I_0}{|g_{11}|} =
\frac{\sqrt{\pi}I_0}{2\sqrt{2}}\frac{a_{1,z}}{|a_1|} =
\kappa_{2D}\frac{a_{1,z}}{|a_1|}, \label{Eq9}
\end{equation}
where $\kappa_{2D} = 1.167$. In the derivation of formula (\ref{Eq9}) we have
used the well-known condition for collapse in the critical nonlinear
Schr\"odinger equation (for details consult Ref.~\cite{Collapse2D}) and that
the number of particles $N_0$ in the so-called Townes soliton is $N_0 = 2\pi
I_0$, where $I_0 = 1.862$.

It is important to notice that both the upper bound (\ref{Eq8}) on the
admissible numbers of atoms and the threshold number for collapse (\ref{Eq9})
in the ${}^{85}$Rb condensate are proportional to the trap size in the
$z$-direction. Thus, taking a bigger pancake trap (with the same $\gamma$)
will relax the bounds on the numbers of atoms. The threshold for collapse in
the mixture of the two isotopes of Rubidium also depends on the number of
atoms of the ${}^{87}$Rb isotope. However, we have found numerically that
this dependence is  very weak (the correction does not exceed 5\% for the
numbers of atoms used below and the fixed default scattering lengths),
therefore, the threshold given by (\ref{Eq9}) can be taken as a good
approximation. For example, for the pancake trap with $a_z = 10\mu$m, used
above, the threshold for collapse is $N_c = 535$.

Finally, let us comment on the validity of the approximate 2D
system for description of the collapse instability in the mixture.
First of all, one may point out that the threshold value for
collapse of the mixture in the pancake trap determined from the
full three-dimensional system (\ref{Eq1ab}) may turn out to be
lower than that predicted by the two-dimensional approximation, as
it is true, for instance, for the single species condensate of
${}^{85}$Rb. Indeed, in the latter case, the exact (i.e., 3D)
threshold can be written as $N_c=\kappa(\gamma){a_{z}}/{|a_1|}\;$
\cite{Collapse3D}. Using the numerically found values of
$\kappa(\gamma)$ from Ref.~\cite{Collapse3D}, we conclude that
$\kappa(\gamma) < \kappa_{2D}$ for any $\gamma>1$, i.e., this
inequality holds for any pancake trap. As $\gamma\to\infty$, the
function $\kappa(\gamma)$ slowly tends to $\kappa_{2D}$. For
example, for the trap with $a_z = 10\mu$m and $\gamma = 100$ we
have $\kappa(100) = 1.1$ \cite{Collapse3D} what gives 95\% (506
atoms) of the threshold given by formula (\ref{Eq9}).

Nevertheless, in the pancake trap, the instability which is solely due to the
three-dimensionality is weak  if the conditions given by (\ref{Eq8}) are
satisfied and the numbers of atoms are not much greater than the
corresponding instability threshold. This conclusion follows from the general
discussion of the 2D approximation, which is placed in
appendix~\ref{appendA}. Here we note also that the instability rate due to
the 3D effects decreases with increase of the trap anisotropy $\gamma$ (since
it enters the r.h.s.'s of the conditions in equation (\ref{Eq8})). Therefore,
in a sufficiently anisotropic pancake trap, the 3D collapse instability below
the threshold of the 2D collapse does not have enough time to develop on the
time scale set by the two-dimensional system (\ref{Eq6ab}) and, hence, its
effect on the solutions can be neglected. In fact, the time necessary for
such a weak instability to develop may exceed the life-time of the
condensates in the mixture.

It is, however,  the {\it dynamics\/} of a collapsing condensate in the
pancake trap that cannot be treated in the framework of the two-dimensional
approximation and requires the full 3D analysis due to violation of at least
one of the two conditions (\ref{Eq8}). Thus we will not discuss such
dynamics. For more details on the two-species collapse in BEC consult, for
instance, Ref.~\cite{multicoll} and on the collapse in boson-fermion mixtures
consult Refs.~\cite{CollapseBF,FullBF}.

In the case of the two-species BEC of ${}^{85}$Rb and ${}^{87}$Rb, there is a
stable state in the mixture, predicted by the 2D system (\ref{Eq6ab}) for the
numbers of atoms slightly lower then the collapse instability (see the next
section), which violates the applicability conditions (\ref{Eq8}) for modest
pancake traps ($\gamma\le 100$) due to sharp contraction of the ${}^{85}$Rb
condensate. The sharp decrease of the ${}^{85}$Rb condensate size $R_1$,
predicted by system (\ref{Eq6ab}), requires large trap anisotropy $\gamma$
for the 2D system to sustain its validity. Therefore, for the current
experimental traps, the very existence of such exotic states requires the
full 3D analysis and thus is beyond the 2D approximation adopted in the
present paper. We will not discuss such states either.

Therefore, for the current experimental pancake traps, the applicability of
the approximate 2D system (\ref{Eq6ab}) is limited by the threshold of
formation of the contracted states in the ${}^{85}$Rb condensate. In the next
section we discuss the ground states in the mixture for the allowed numbers
of atoms and their deformations due to the instabilities predicted by the 2D
system, such as the symmetry-breaking transition. Such instabilities are much
stronger than those due to the three-dimensional effects and, consequently,
are observed on much shorter time scale (consult also
appendix~\ref{appendA}).

\section{Ground states in the mixture of two isotopes of Rubidium}
\label{III}

Now we turn to the numerical solution of system (\ref{Eq6ab}) to find
possible ground states in the BEC mixture of the two isotopes. The stationary
solutions are sought for in the usual form:
\begin{equation}
\psi_1 = e^{-i\mu_1 T} U_1(\rho), \quad \psi_2 = e^{-i\mu_2 T}
U_2(\rho), \label{Eq10}\end{equation}
 where $\mu_1$ and $\mu_2$ are the dimensionless chemical
 potentials for the two species.  We have used the gradient method for the constrained
energy minimization to find $U_1(\rho)$ and $U_2(\rho)$ minimizing  the
energy functional

\begin{equation}
\mathcal{E} = \int\mathrm{d}^2\bm{\rho}\left\{ |\nabla_\perp\psi_1|^2 +
|\nabla_\perp\psi_2|^2 + \rho^2(\lambda_1^2|\psi_1|^2 +
\lambda^2_2|\psi_2|^2) + \frac{g_{11}}{2}|\psi_1|^4 +
\frac{g_{22}}{2}|\psi_2|^4 + g_{12}|\psi_1\psi_2|^2\right\}
\label{Eq11}\end{equation}
 for fixed numbers of atoms $N_1=\int\mathrm{d}^2\bm{\rho}|\psi_1|^2$ and $N_2 =
\int\mathrm{d}^2\bm{\rho}|\psi_2|^2$.

\subsection{Ground states for repulsive interspecies interaction}
\label{IIIa}

Let us start with considering the  BEC mixture of ${}^{85}$Rb and ${}^{87}$Rb
atoms with the repulsive interspecies interaction. First of all, we have
found the axially symmetric ground states via the energy minimization
restricted to the space of the axially symmetric functions. It is important
to know if the symmetric states are stable. The stability analysis cab be
based on the method of Refs.~\cite{Kuzn,Peli}, whose adaptation to our
problem is described in appendix~\ref{appendB}. We have found that the
axially symmetric ground  state of the mixture suffers from the dipole-mode
symmetry-breaking instability  for sufficiently large number of atoms in the
${}^{87}$Rb condensate and not too large numbers of atoms in the ${}^{85}$Rb
condensate ($N_{85} \le 500$). The symmetry breaking instability was
previously discussed for the case of BEC mixtures in
Refs.~\cite{symbr1,symbr2,symm_break,symm_br_pan,sym_asym_trans}. The novelty
here lies in the fact that one of the condensates has attractive atomic
interaction. For instance, the separation criterion of
Ref.~\cite{crit_repuls} for a BEC mixture in the box, i.e.,
$g_{12}>\sqrt{g_{11}g_{22}}$, looses its meaning since in our case
$g_{11}g_{22} < 0$ and,  {\it a priori}, it is not evident that the isotopes
would separate at all.

\begin{figure}
\includegraphics{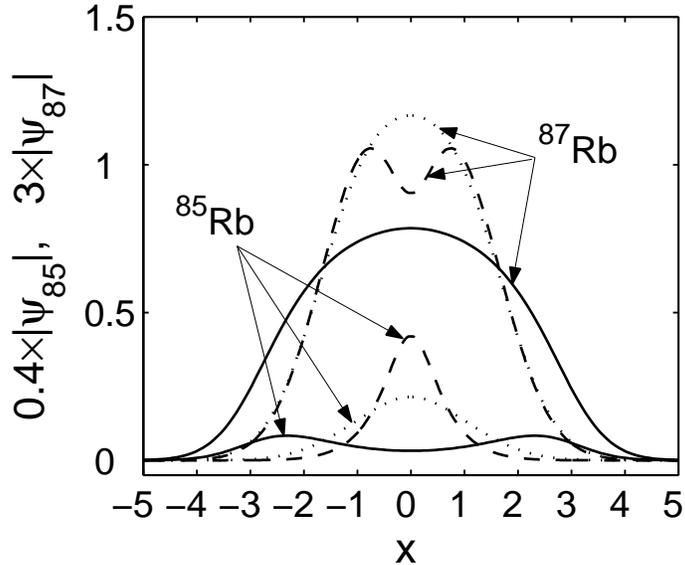}
\caption{\label{F1} The three types  of the axially symmetric state in the
mixture of ${}^{85}$Rb and ${}^{87}$Rb isotopes on the threshold of the
symmetry-breaking instability. The one-particle wave functions are shown
(scaled for better visibility as  is indicated on the $y$-axis). The numbers
of atoms are as follows. Solid lines: $N_{85} = 100$ and $N_{87} = 17412$;
dotted lines: $N_{85} = 300$ and $N_{87} = 2042$; and dashed lines: $N_{85} =
450$ and $N_{87} = 894$. }
\end{figure}

The axially symmetric ground states  on the threshold of the
symmetry-breaking  instability for various numbers of atoms are shown in
figure~\ref{F1}. It should be stressed that there are three types of the {\it
stable} axially symmetric states in the system for smaller numbers of atoms,
which correspond (and are similar) to the threshold states shown in
figure~\ref{F1}: ($i$) when the isotopes are strongly mixed (the two
condensates have bell-shaped form, the doted lines), ($ii$) when the
${}^{85}$Rb isotope is on the surface of ${}^{87}$Rb ($|\psi_1|$ has a local
minimum at the center, the solid lines), and ($iii$) when the ${}^{87}$Rb
isotope is on  the surface of ${}^{85}$Rb isotope ($|\psi_2|$ has a local
minimum at the center, the dashed lines).

\begin{figure}
\includegraphics{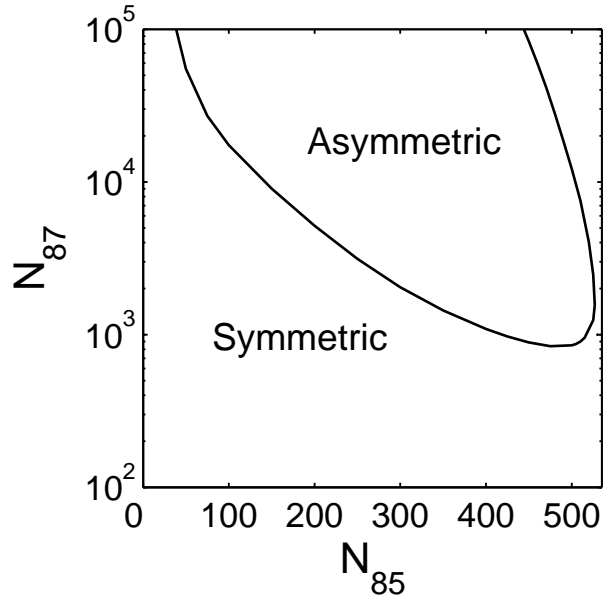}
\caption{\label{F2} The symmetric vs.  asymmetric ground state diagram. The
interaction coefficients are: $g_{11} = -0.0219$, $g_{22} = 0.0068$, and
$g_{12} = 0.012$ (computed for the default values of the scattering lengths
and $a_{1,z} = 10$ $\mu$m). The logarithmic (base 10) scale is used for the
${}^{87}$Rb-axis. }
\end{figure}

The threshold of the symmetry-breaking  instability strongly depends on the
numbers of atoms and corresponds to appearance of a zero dipole mode (which
pertains to the orbital operator $\Lambda_{11}$, consult
appendix~\ref{appendB}). From the energetic point of view, the separation
takes place when the energy gain due to the intraspecies interaction in the
strongly mixed state is higher than the kinetic energy (quantum pressure) at
the interface between the separated condensates. The corresponding
symmetry-breaking diagram, found numerically, is given in figure~\ref{F2}.

Though all three types of the axially  symmetric ground states discussed
above suffer from the symmetry-breaking instability with increase of the
number of atoms of the ${}^{85}$Rb isotope (for sufficiently large number of
atoms of the ${}^{87}$Rb isotope) the symmetry-restored states (found to the
right of the phase separation curve in figure~\ref{F2}), which result from
further increase of the number of ${}^{85}$Rb atoms, are of type $(iii)$,
i.e., when the ${}^{87}$Rb isotope is on the surface of the ${}^{85}$Rb
isotope.

In the reduced 2D system (\ref{Eq6ab}),  with further increase of the number
of ${}^{85}$Rb atoms the symmetry-restored state of type $(iii)$ is
immediately followed by a sharp contraction of the ${}^{85}$Rb condensate and
subsequent collapse instability at $N_{85} \approx 535$. The collapse
instability is due to appearance of the axially symmetric unstable mode
(i.e., the unstable linear mode with the orbital number \mbox{$l=0$}, see
appendix~\ref{appendB}). The collapse threshold value of $N_{85}$ only
slightly decreases with increase of the number of ${}^{87}$Rb atoms. This is
due to the very favorable set of the default scattering lengths of the system
and is not a general property of the mutually-repulsive mixtures of
attractive and repulsive species. For instance, in the related study of the
boson-fermion mixtures, it was noted that though the collapse in the
mutually-repulsive boson-fermion mixture with attractive boson interactions
only concerns the boson species, it can be strongly affected by the fermion
number of atoms \cite{FullBF}. However, for such effect to be pronounced, the
interspecies $s$-wave scattering length must be significantly larger than the
absolute value of the boson scattering length.

Thus, right before the collapse instability   the size of the ${}^{85}$Rb
condensate first decreases to a fraction of the trap size $a_\perp$. However,
depending on the trap anisotropy $\gamma$, such a state may violate the first
of the two applicability conditions (\ref{Eq8}) for the 2D approximation. For
example,  our estimates show that accurate description of this effect
requires the full 3D analysis for the pancake traps with $\gamma \le 100$.
For observation of this effect, much more anisotropic pancake traps are
required, which are not used in the current experiments. Thus we will not
discuss the effect any further. There is also an implication on the validity
of a part of figure~\ref{F2} for the current pancake traps: the 2D
approximation for the pancake trap with $\gamma\le100$ is not valid for
description of the symmetric ground states to the right of the separation
curve except a narrow strip immediately after it (with the width equal to a
dozen of atoms on the $N_{85}$ axis).

\begin{figure}
\includegraphics{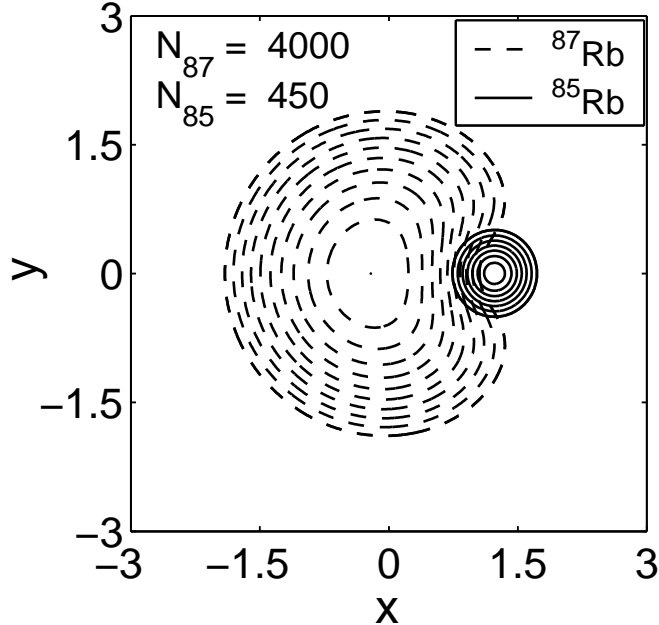}
\caption{\label{F3} The symmetry-breaking ground state for not too
large number of ${}^{87}$Rb atoms. For each of the two
condensates, the equidistant level curves ranging from the maximum
of  the order parameter to a half of its value are shown. }
\end{figure}

\begin{figure}
\includegraphics{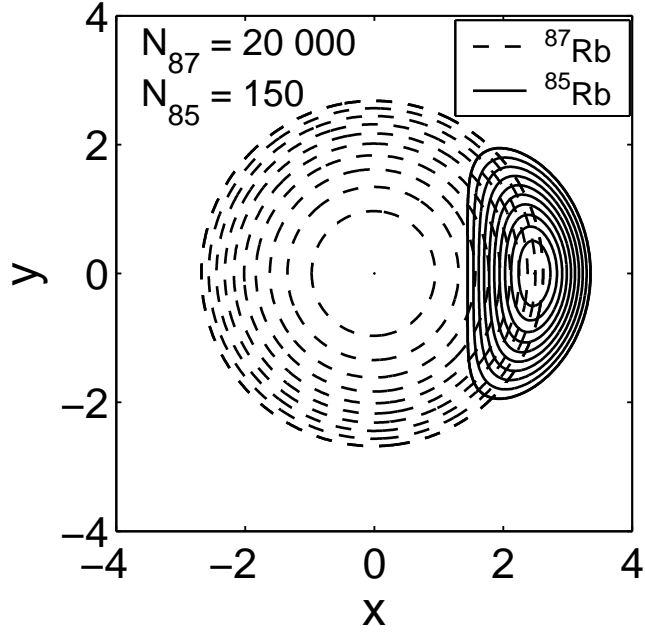}
\caption{\label{F4} The asymmetric ground state for very large
number of ${}^{87}$Rb atoms. The equidistant level curves are
shown with the range defined  as in the previous figure. }
\end{figure}

The symmetry-breaking  ground states are illustrated in figures~\ref{F3}
and~\ref{F4}, where we show the contour lines of the order parameters
(ranging from the maximum to half of its value at a constant step) for
${}^{85}$Rb (solid lines) and ${}^{87}$Rb (dashed lines). We have found that
it is the ${}^{85}$Rb condensate that is expelled from the center of the trap
in the symmetry-breaking states. It is seen that for comparable numbers of
atoms of the two species it is the ${}^{87}$Rb condensate that suffers the
strongest deformation from the bell-shaped form, while for $N_{87}\gg N_{85}$
the ${}^{85}$Rb condensate is strongly deformed. Here we note that the
asymmetric ground state of the mixture is degenerate, as it possesses the
rotational zero mode. In other words, the maximum of the order parameter of
${}^{85}$Rb can have arbitrary position angle on the surface of the
${}^{87}$Rb condensate.

\subsection{Ground states for attractive interspecies interaction}
\label{IIIb}

\begin{figure}
\includegraphics{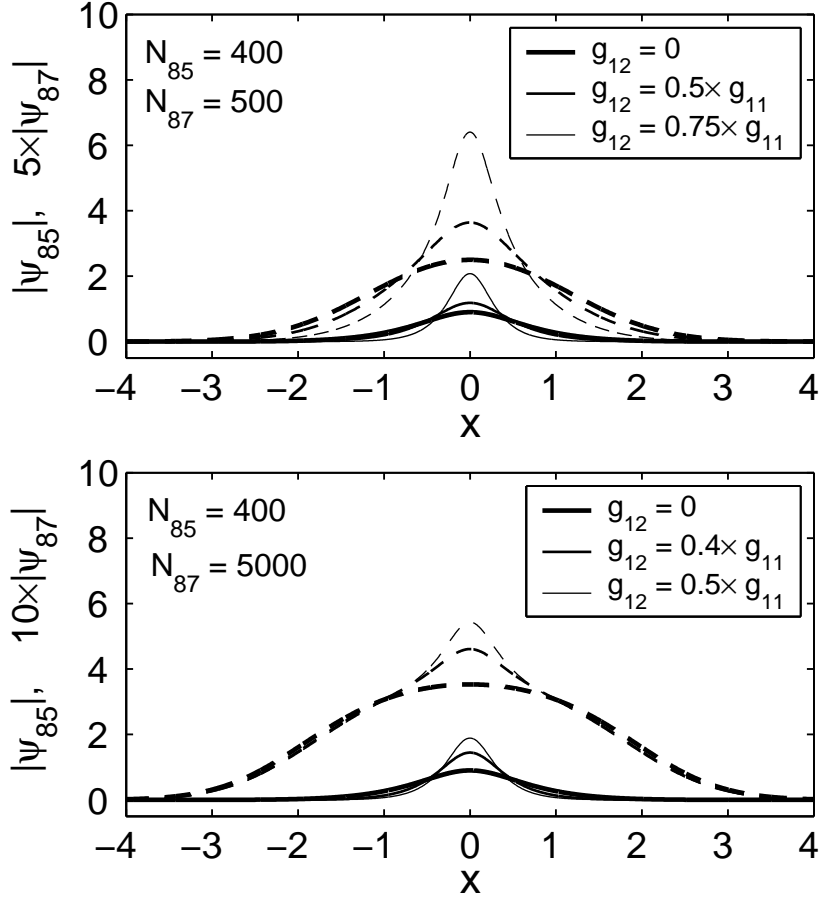}
\caption{\label{F5} Stable  ground states of the BEC mixture for the
attractive interspecies interactions. The one-particle wave functions (scaled
for better visibility as indicated along the $y$-axis) are shown. In both
panels, the solid lines correspond to the ${}^{85}$Rb isotope and the dashed
lines to the ${}^{87}$Rb isotope. }
\end{figure}

Now let us consider the BEC mixture of the two isotopes when the interspecies
interaction is attractive, which can be experimentally realized by using the
Feshbach resonance \cite{prospects}.  It is convenient to measure the
interspecies interaction coefficient $g_{12}$ in terms of the interaction
coefficient $g_{11}$ of the ${}^{85}$Rb isotope. We have found that the
condensates do not separate in this case and assume the axially symmetric
ground state up to the collapse instability threshold. Such ground states are
illustrated in figure~\ref{F5}, where we plot the appropriately scaled
one-particle wave-functions for the two condensates. Note the local peak at
the center of the ${}^{87}$Rb condensate. Appearance of this peak is easy to
understand, for instance, when the Thomas-Fermi limit applies to the
${}^{87}$Rb condensate (which corresponds to the picture in the lower panel
of figure~\ref{F5}). Indeed, if $N_{87}\gg N_{85}$, then in the zero-order
approximation one can neglect the contribution from the interspecies
interaction term $g_{12}|\psi_1|^2$  in equation (\ref{Eq6b}) for the
${}^{87}$Rb condensate in the region away from the trap center. Thus, in the
zero-order approximation, the ${}^{87}$Rb condensate has the Thomas-Fermi
ground state independently of the state of the other isotope. Therefore the
effect of the cross-interaction term $g_{12}|\psi_2|^2$ in the equation  for
the ${}^{85}$Rb condensate (\ref{Eq6a})  is now similar to that of an
additional potential. In the next order of approximation, the order parameter
$\psi_2$ of the ${}^{87}$Rb condensate  is a sum of the two terms: the
background Thomas-Fermi shape and the local deformation at the center of the
trap. The latter is determined by the order parameter of the ${}^{85}$Rb
condensate. It is interesting to note that a similar central density
enhancement of the fermion species in the boson-fermion mixtures with
attractive interspecies interactions is predicted in Ref.~\cite{FullBF}
(figures 1 and 2).

It should be noted that,  notwithstanding the attraction in the ${}^{85}$Rb
condensate and the attractive interspecies interaction, the ground states
shown in figure~\ref{F5} are not self-bound, i.e., they are not solitons:
relaxation of the trap results in spreading of the condensates.

Finally, lowering  of the   interspecies interaction coefficient $g_{12}$ to
the sufficiently large negative values results in sharp contraction of the
${}^{85}$Rb condensate which is followed by the collapse instability.  This
contraction of the {\it stable} ground state of the ${}^{85}$Rb condensate to
a fraction of the trap size is due to the presence of the other condensate,
since the size of a stable single species BEC is always on the order of the
trap size.

Thus, as in the case of the  repulsive interspecies interactions, the
collapse instability in the mixture is preceded by a sharp decrease of the
${}^{85}$Rb condensate size. Therefore, depending on the trap anisotropy
$\gamma$, its description may take us beyond the 2D approximation adopted in
this paper. For most of the current pancake traps $\gamma\le 100$, thus the
description of the ground state in the mixture which is on the border of the
collapse instability requires the full three-dimensional analysis.

We note, however, that for  small values of $|g_{12}|$ the stable symmetric
ground states  predicted by the 2D approximation and illustrated in figure
\ref{F5} can be experimentally observed in the current pancake traps with
$\gamma \sim 100$. In such experiment, the number of ${}^{85}$Rb atoms should
be below the threshold value $N_c = N_c(N_{87})$ which, for fixed $g_{12}$,
decreases only by a few percent from the value given by formula (\ref{Eq9})
with increase of the number of ${}^{87}$Rb atoms.

\section{Conclusion}
\label{IV}

We have studied the ground state in  the BEC mixture of two isotopes of
Rubidium in the pancake trap for repulsive and attractive interspecies
interactions and fixed (default) intraspecies interactions.

In the case of repulsion between the  two species, there is the
symmetry-breaking deformation of the ground state due to the dipole-mode
instability, whose threshold strongly depends on the numbers of atoms of the
two isotopes. For small numbers of atoms, i.e., below the symmetry-breaking
instability threshold, the stable axially symmetric ground state has the form
of either the strongly mixed state of the species (with the order parameters
of the two condensates having the bell-shaped form) or the state where one of
the condensates forms a circular strip on the surface of the other.

For attractive interspecies interaction, the condensates assume the axially
symmetric ground states  for all numbers of atoms where the 2D approximation
is valid. For small values of the (negative) interspecies interaction
coefficient $|g_{12}|$ the mixture is stable for the numbers of ${}^{85}$Rb
atoms below the critical value $N_c = N_c(N_{87})$ which is a few percent
lower then the collapse instability threshold for a single species condensate
of ${}^{85}$Rb (i.e., for $g_{12}=0$). There is a sharp peak in the density
of the repulsive ${}^{87}$Rb isotope due to the attractive interspecies
interactions -- the effect which is similar to the  enhancement of the
fermionic density in boson-fermion mixtures with attractive boson-fermion
interactions \cite{FullBF}.

Finally, for the pancake traps   with the anisotropy $\gamma\le 100$, the 2D
approximation for the attractive as well as repulsive mixture is violated at
the numbers of atoms in the ${}^{85}$Rb condensate just below the collapse
instability in the 2D model due to the sharp contraction of the ${}^{85}$Rb
condensate. This is quite dissimilar to the case of the single species BEC of
${}^{85}$Rb, where the collapse sets in at a state which has the size
comparable to the trap size in the pancake plane. Thus, the investigation of
the actual ground state of the mixture in such pancake traps at the numbers
of atoms close to the collapse instability requires the full
three-dimensional analysis and is beyond the approach adopted in the present
paper. This sets a direction for the future research.

\section*{Acknowledgements}
This work was supported by the FAPESP of Brazil. A.M.K. would like
to thank the Instituto de F\'{i}sica Te\'{o}rica - UNESP for kind
hospitality. We are grateful to the referee for the suggestions
which resulted in significant improvement of the presentation.

\appendix

\section{Derivation of the two-dimensional approximation}
\label{appendA}

The system (\ref{Eq1ab}) can be rewritten in the following form:
\begin{eqnarray}
i\hbar \partial_t \Psi_1 &=& \left( H_{1z} + H_{1\perp}
+ G_{11}|\Psi_1|^2 + G_{12}|\Psi_2|^2 \right)\Psi_1, \label{B1a}\\
i\hbar \partial_t \Psi_2 &=&  \left(H_{2z} + H_{2\perp} +
G_{22}|\Psi_2|^2 + G_{12}|\Psi_1|^2\right) \Psi_2, \label{B1b}
\end{eqnarray}
where we have introduced the linear operators corresponding to the
quantum motion along $z$-axis and on the $(x,y)$-plane in the
trap:
\[
H_{jz} = -\frac{\hbar^2}{2m_j} \frac{\partial^2}{\partial z^2} +
V_{jz}(z),\quad H_{j\perp} = -\frac{\hbar^2}{2m_j} \nabla_\perp^2
+ V_{j\perp}(\bm{r}_\perp),\quad j=1,2.
\]

The solution  to equations (\ref{B1a})-(\ref{B1b}) can be expanded
over the eigenfunctions of the linear operators $H_{1z}$ and
$H_{2z}$ as follows: $\Psi_j = \Psi_{j0}(z,\bm{r}_\perp,t) +
\hat{\Psi}_{j}(z,\bm{r}_\perp,t)$, $j=1,2$, where $\Psi_{j0}\equiv
e^{-iE_{j0}t/\hbar}f_j(z)\Phi_j(\bm{r}_\perp,t)$, with $f_j(z)$
being the normalized eigenfunction of the ground state,
$H_{jz}f_j(z) = E_{j0}f_j(z)$, while the second term,
$\hat{\Psi}_j$, is the projection of order parameter $\Psi_j$ on
the subspace orthogonal to the ground state.

We can expand the system (\ref{B1a})-(\ref{B1b}) in similar way
using the projectors $\Pi_1$ and $\Pi_2$ on the ground states of
$H_{1z}$ and $H_{2z}$, respectively. Application of these
projectors  to equations (\ref{B1a})-(\ref{B1b}) leads to the
system describing evolution of the projection of the order
parameters for the two species on the respective ground states:
\begin{eqnarray}
i\hbar \partial_t \Phi_1 &=& \left(  H_{1\perp} +
\tilde{G}_{11}|\Phi_1|^2 + \tilde{G}_{12}|\Phi_2|^2  +
\Delta_{1}\right)\Phi_1  + \mathcal{F}_1,
\label{B3a}\\
i\hbar \partial_t \Phi_2 &=& \left( H_{2\perp} +
\tilde{G}_{22}|\Phi_2|^2 + \tilde{G}_{12}|\Phi_1|^2 +
\Delta_{2}\right)\Phi_2 + \mathcal{F}_2, \label{B3b}
\end{eqnarray}
where $\tilde{G}_{ij} = G_{ij}\langle f_i^2f_j^2 \rangle$, with
$\langle F \rangle \equiv \int F\mathrm{d}z$,
\begin{equation}
\Delta_{j} = G_{jj}\langle
f^2_j(2f_i\mathrm{Re}\{\Phi_j\hat{\Psi}^*_j\}+|\hat{\Psi}_j|^2)\rangle
+ G_{j,3-j}\langle
f_j^2(2f_{3-j}\mathrm{Re}\{\Phi_{3-j}\hat{\Psi}^*_{3-j}\} +
|\hat{\Psi}_{3-j}|^2)\rangle, \label{DELTs}\end{equation}
 and
\begin{equation}
\mathcal{F}_j = G_{jj}\langle f_j|\Psi_j|^2\hat{\Psi}_j\rangle +
G_{j,3-j}\langle f_j|\Psi_{3-j}|^2\hat{\Psi}_j\rangle.
\label{Fs}\end{equation}
 On the other hand, the equations
describing evolution of the projections of the order parameters on
the orthogonal subspaces are derived by application of the
complementary projectors $Q_j$, $Q_j \equiv\openone - \Pi_j$, to
the system (\ref{B1a})-(\ref{B1b}):
\begin{eqnarray}
i\hbar \partial_t \hat{\Psi}_1 &=& \left( \tilde{H}_{1z} +
H_{1\perp}\right)\hat{\Psi}_1 + Q_1\left\{G_{11}|\Psi_1|^2\Psi_1+
G_{12}|\Psi_2|^2\Psi_1\right\},
\label{B8a}\\
i\hbar \partial_t \hat{\Psi}_2 &=& \left( \tilde{H}_{2z} +
H_{2\perp} \right)\hat{\Psi}_2 + Q_2\left\{G_{22}|\Psi_2|^2\Psi_2
+ G_{12}|\Psi_1|^2 \Psi_2\right\}, \label{B8b}
\end{eqnarray}
where $\tilde{H}_{jz} = H_{jz} - E_{j0}$.

Let us estimate the orders of magnitude of the nonlinear terms in
equations (\ref{B3a})-(\ref{B3b}) and (\ref{B8a})-(\ref{B8b})
under the condition that the nonlinear terms in the system
(\ref{B1a})-(\ref{B1b}) are much smaller  than the characteristic
difference $\Delta E\sim\hbar\omega_z$ between the eigenvalues  of
each of the two linear operators $H_{1z}$ and $H_{2z}$ (i.e., the
conditions (\ref{Eq3}) are satisfied). Below we will not
distinguish between the quantities with different indexes, since
all  quantities of the same kind are of the same order of
magnitude. Introduce a small parameter $\epsilon$ as the ratio of
the nonlinear terms in the system (\ref{B3a})-(\ref{B3b}) to
$\Delta E$, that is
\begin{equation}\label{small_param}
\epsilon=\frac{\tilde{G}|\Phi|^2}{\Delta E}=\frac{G\langle
f^4\rangle|\Phi|^2}{\Delta E}.
\end{equation}
Note that $\langle f^4\rangle \sim f^2$ since the integral of
$f^2$ over $z$ is of order 1. It is the smallness of $\epsilon$,
supposed in (\ref{Eq3}), that justifies the transition to the
two-dimensional approximation. Indeed, the order of the correction
$\hat{\Psi}$ to the factorized wave function can be found by
equating the orders of the inhomogeneous term and the linear term
$H_z + H_\perp$ in (\ref{B8a})-(\ref{B8b}), where, as in
(\ref{B3a})-(\ref{B3b}), we have again $H_\perp\hat{\Psi} \sim
\epsilon\Delta E\hat{\Psi}$ and $H_\perp$ can be neglected
compared with $H_z\sim\Delta E$. Therefore, we get $\Delta
E\hat{\Psi}\sim G Q\{|f\Phi|^2f\Phi \} \sim \epsilon \Delta
Ef\Phi$ and, hence, $\hat{\Psi} \sim \epsilon f\Phi$. From this we
obtain the estimates for the terms given by equations
(\ref{DELTs}) and (\ref{Fs}): $\Delta \sim G \langle
f^2f\Phi\hat{\Psi}\rangle \sim \epsilon Gf^2\Phi^2 \sim
\epsilon^2\Delta E$ and $\mathcal{F} \sim G\langle f
(f\Phi)^2\epsilon f\Phi\rangle \sim \epsilon G f^2\Phi^2\Phi \sim
\epsilon^2 \Phi\Delta E$.

Throwing away the terms of the order $\epsilon^2$ from equations
(\ref{B3a})-(\ref{B3b})  and changing to the dimensionless
variables given by (\ref{Eq5}) we arrive at the system
(\ref{Eq6ab}). The projection of the order parameter $\Psi_j$ on
the orthogonal subspace, $\hat{\Psi}_j$, is of order $\epsilon$
and can be neglected compared to the factorized wave function
$f_j\Phi_j$. We conclude that, under the conditions (\ref{Eq3}),
nonlinearity plays significant role only on the pancake plane
$(x,y)$.

Two comments are in order on the two-dimensional approximation
described above. First, the effects due to three-dimensionality of
the mixture will be of order $\epsilon^2$, the same order as the
terms we have thrown out from the system (\ref{B3a})-(\ref{B3b}),
thus they will give a significant contribution to the dynamics
only on the time scale of order $\epsilon^{-2}$, much longer then
the time scale of the nonlinear effects in 2D, which is of order
$\epsilon^{-1}$. Second, a similar 2D approximation will be valid
for the linearized system which describes the evolution of a small
perturbation of the solution. Thus, any instability in the mixture
which is solely due to its three-dimensionality will be of order
$\epsilon^2$ and will not play any role on the time scale we
consider.

The latter comment concerns, for instance, the instability due to
collapse in 3D: although the 3D threshold value of the number of
atoms in the mixture necessary for collapse may turn out to be
lower than that in the 2D approximation, as it is true for the
single species condensate of ${}^{85}$Rb, the corresponding
instability rate (proportional to the unstable eigenvalue) will be
of order $\epsilon^2$ and will not be noticed on the time scale we
consider.

\section{The Linear stability analysis}
\label{appendB}

 The linear stability analysis is based on the
consideration of  evolution of a linear perturbation $u_1 =
u_1(\bm{\rho},T)$ and  $u_2 = u_2(\bm{\rho},T)$ of the stationary
state $(U_1(\rho),U_2(\rho))$. The evolution equations for the
perturbation are derived by linearization of the original system
(system (\ref{Eq6ab})) about the stationary solution. One looks
for the eigenfrequencies $\omega$ of the resulting linear system
by setting $u_j = e^{-i\omega t}(X_j(\bm{\rho}) +
iY_j(\bm{\rho}))$, $j=1,2$. Appearance of an imaginary
eigenfrequency means instability. In particular, by writing the
perturbed solution as
\begin{equation}
\psi_j = e^{-i\mu_j T}(U_j(\rho) + u_j(\bm{\rho},T)), \quad j=1,2,
\label{A1}\end{equation}
 using this in the system (\ref{Eq6ab}) and keeping  only the linear
terms in $u_1$ and $u_2$ we arrive at the following eigenvalue
problem (consult also Refs.~\cite{Kuzn,Peli}):
\begin{equation}
\Lambda_0\left(\begin{array}{c}Y_1\\ Y_2\end{array}\right) =
-i\omega\left(\begin{array}{c}X_1\\ X_2\end{array}\right), \quad
\Lambda_1\left(\begin{array}{c}X_1\\ X_2\end{array}\right) =
i\omega\left(\begin{array}{c}Y_1\\ Y_2\end{array}\right),\quad \Lambda_n =
\left(\begin{array}{cc}L_{n1} & 0\\ 0 & L_{n2}\end{array}\right).
\label{A2}\end{equation} Here ($j=1,2$):
\begin{eqnarray}
L_{0j} &=& -\mu_j - \nabla^2_\perp + \sum_{m=1,2}g_{jm}U^2_m(\rho)
+ \lambda_j^2 \rho^2,\nonumber\\
& & \label{A3}\\
L_{1j} &=& -\mu_j - \nabla^2_\perp + \sum_{m=1,2}g_{jm}U^2_m(\rho) +
2g_{jj}U^2_{j}(\rho) + \lambda_j^2 \rho^2.
\nonumber\end{eqnarray}

Expansion of the eigenvalue problem (\ref{A2}) in the Fourier
series with respect to the polar angle $\theta$ leads to an
infinite series of one-dimensional eigenvalue problems of similar
form for the orbital projections of the vectors $(X_1,X_2)^T$ and
$(Y_1,Y_2)^T$,
\begin{equation}
X_j  = \sum_{l\ge0} X_{jl}(\rho)e^{il\theta} +\mathrm{c.c.},\quad Y_j  =
\sum_{l\ge0} Y_{jl}(\rho)e^{il\theta} +\mathrm{c.c.},\quad j=1,2,
\label{Eq14}
\end{equation}
with the orbital operators defined by {$\Lambda_{0l} =
\Lambda_0(\nabla^2_\perp\to\nabla^2_\rho - l^2/\rho^2)$ and
$\Lambda_{1l} = \Lambda_1(\nabla^2_\perp\to\nabla^2_\rho -
l^2/\rho^2)$, where $\nabla^2_\rho\equiv \partial_\rho^2 +
\rho^{-1}\partial_\rho$. However, only a few of these 1D
eigenvalue problems need to be considered to decide on stability
of the axial ground state. Indeed, first, as follows from the
general criterion for stability of the ground state in a system of
nonlinear Schr\"odinger equations \cite{Peli}, in the
two-component system the axial ground state is unstable if there
are at least three negative eigenvalues of the operator
$\Lambda_1$. For instance, if each of the first three orbital
operators $\Lambda_{1l}$, $l=0,1,2$ has a negative eigenvalue,
then the ground state is unstable. Second, as the orbital
operators satisfy the obvious inequality $\Lambda_{1l_2} \ge
\Lambda_{1l_1}$ (understood as the inequality for the mean values)
for $l_2 \ge l_1$,  it is sufficient to consider just three
orbital problems arising from (\ref{A2}) with the orbital numbers
$l=0,1,2$. This is the approach we adopted.

Finally, we would like to mention that the linear stability is
closely related to the energy minimization (consult also
Refs.~\cite{Kuzn,Peli}). Indeed, the operator $\Lambda_0$ is
non-negative (it has two zero modes due to the phase invariance of
the system (\ref{Eq6ab})). Thus, from the energetic point of view,
a negative eigenvalue of the operator $\Lambda_1$ corresponds to a
negative direction in the free energy functional, defined here as
the Lagrange-modified energy functional: $\mathcal{F} \equiv
\mathcal{E} - \mu_1N_1 - \mu_2N_2$, evaluated at the axially
symmetric state $(U_1(\rho),U_2(\rho))$, since the operator
$\Lambda_1$ enters the second-order term in the free energy
expansion with respect to the perturbation $\hat{u}_j =
X_j(\bm{\rho})+iY_j(\bm{\rho})$, $j=1,2$:
\begin{equation}
\delta^2\mathcal{F} =
2\int\mathrm{d}^2\bm{\rho}\left\{(Y_1,Y_2)\Lambda_0\left(\begin{array}{c}Y_1 \\
Y_2\end{array}\right) + (X_1,X_2)\Lambda_1\left(\begin{array}{c}X_1 \\
X_2\end{array}\right)\right\}.
 \label{A4}\end{equation}
Taking into account that there are two independent constraints on the
numbers of atoms in the two species  we conclude that only two negative
directions may be eliminated by the energy dependence on the numbers of
atoms. Therefore, for fixed numbers of atoms, the axially symmetric state
is definitely not an energy minimum if there are three (or more) negative
eigenvalues of the operator $\Lambda_1$. This explains the physical origin
of the above mentioned sufficient condition for instability. We note also
that the dipole-mode instability (i.e., existence of the unstable orbital
mode with the orbital number $l=1$) simply means appearance of an
asymmetric state which minimizes the energy, i.e., the symmetry-breaking
transition.

\end{document}